\documentclass[%
 aip,
 amsmath,amssymb,
 reprint,%
]{revtex4-2}

\usepackage{graphicx}
\usepackage{dcolumn}
\usepackage{bm}
\usepackage{layouts}
\usepackage[utf8]{inputenc}
\usepackage[T1]{fontenc}
\usepackage{mathptmx}
\usepackage{color}

\begin{document}

\preprint{AIP/123-QED}


\title{A Robust Coherent Single-Photon Interface for Moderate- NA Optics Based on SiV Center in Nanodiamonds  and a Plasmonic Bullseye Antenna.}

\author{R. Waltrich}
 \altaffiliation{Institute for Quantum Optics, Ulm University, Albert Einstein-Alle 11, 89081 Ulm, Germany}
 
\author{H. Abudayyeh}
\altaffiliation{Racah Institute of Physics and the Center for Nanoscience and Nanotechnology, The Hebrew University of Jerusalem, Jerusalem 9190401, Israel}

\author{B. Lubotzky}
\altaffiliation{Racah Institute of Physics and the Center for Nanoscience and Nanotechnology, The Hebrew University of Jerusalem, Jerusalem 9190401, Israel}

\author{E. S. Steiger}%
\altaffiliation{Institute for Quantum Optics, Ulm University, Albert Einstein-Alle 11, 89081 Ulm, Germany}%

\author{K. G. Fehler}%
\altaffiliation{Institute for Quantum Optics, Ulm University, Albert Einstein-Alle 11, 89081 Ulm, Germany}%

\author{N. Lettner}%
\altaffiliation{Institute for Quantum Optics, Ulm University, Albert Einstein-Alle 11, 89081 Ulm, Germany}%

\author{V. A. Davydov}%
\altaffiliation{4L.F. Vereshchagin Institute for High Pressure Physics, Russian Academy of Sciences, Troitsk, Moscow 142190, Russia}%

\author{V. N. Agafonov}%
\altaffiliation{GREMAN, UMR CNRS CEA 7347, University of Tours, 37200 Tours, France}%

\author{R. Rapaport}
\altaffiliation{Racah Institute of Physics and the Center for Nanoscience and Nanotechnology, The Hebrew University of Jerusalem, Jerusalem 9190401, Israel}

\author{A. Kubanek}
 \altaffiliation{Institute for Quantum Optics, Ulm University, Albert Einstein-Alle 11, 89081 Ulm, Germany}

\date{\today}

\begin{abstract}
Coherent exchange of single photons is at the heart of applied Quantum Optics. The negatively-charged silicon vacancy center in diamond is among most promising sources for coherent single photons. Its large Debye-Waller factor, short lifetime and extraordinary spectral stability is unique in the field of solid-state single photon sources. However, the excitation and detection of individual centers requires high numerical aperture optics which, combined with the need for cryogenic temperatures, puts technical overhead on experimental realizations. Here, we investigate a hybrid quantum photonics platform based on silicon-vacancy center in nanodiamonds and metallic bullseye antenna to realize a coherent single-photon interface that operates efficiently down to low numerical aperture optics with an inherent resistance to misalignment.  
\end{abstract}

\maketitle

\section{\label{sec:Introduction}Introduction\protect\\}

Coherent single photons are a key element for quantum technology such as quantum networks or quantum repeater where coalescence of indistinguishable photons is required to distribute quantum information over distance. A major requirement is the ability to resonantly address a single quantum emitter (SPE) with laser excitation and to efficiently collect coherent single photons. The fundamental challenge arises from the need to cool the solid to cryogenic temperatures and, at the same time, the need for optics with high numerical aperture (NA) which requires short working distance. 

In the past decade there has been considerable efforts to modify the photonic environment near quantum emitter \cite{Dey_2016, Pelton_2015}. To achieve this, emitters were embedded in, or near to various resonant optical structures such as photonic crystal cavities \cite{Fehler_2019} and nano-antennas \cite{Chu_2014}. One approach for improving the directionality and emission rate of quantum emitters is the use of metallic antennas. These include metal nanoparticles \cite{Dey_2016}, plasmonic patch antennas \cite{Esteban_2010, Belacel_2013, Bigourdan_2014, Bogdanov_2018}, metallic nanoslit arrays \cite{Livneh_2011}, Yagi-Uda nanoantennas \cite{Curto_2010,Dregely_2011},  circular bullseye plasmonic nanoantennas \cite{Li_2013,Harats_2014} and plasmonic Bragg cavities \cite{deLeon2012, Siampour2017}. The advantage of using such structures is that plasmonic modes have low mode volumes accompanied with low quality factors enabling spontaneous emission lifetime shortening and emission redirection over broad spectral ranges which can be used to enhance the zero phonon line emission of solid state emitters like the nitrogen vacancy center \cite{deLeon2012, Siampour2017} or recently the germanium vacancy center cite \cite{Kumar2021}. On the other hand, achieving both emission lifetime shortening and high collection efficiency into low-NA optics with pure plasmonic structures require significant plasmon propagation lengths which in turn increase non-radiative recombination rates or quenching of the emission all together \cite{Giannini_2011}.  Another approach is to use pure dielectric antennas such as microcavities \cite{Ding_2016} and photonic crystals \cite{Englund_2009,Laucht_2012,Manga_2007} that feature high radiative enhancement factors and low-loss \cite{Ates_2009,Davanco_2011}. Despite these advantages however, dielectric antennas usually come with a limiting narrow frequency bandwidth and are much more complex to fabricate. 

One solution are hybrid metal-dielectric antenna that combine the advantages of metallic and dielectric antennas but without drawbacks which is typical for metallic antennas, such as increase non-radiative recombination rates or quenching of the emission. In such a design, the emitter can be placed at a large distance from the metal and still produce high directionality in a broad spectral range \cite{Livneh_2016, Livneh_2015}. Recent experiments demonstrated that the single photons emitted from a single photon emitter positioned in such a hybrid circular bullseye antenna can be collected with high efficiency into a moderate numerical aperture. Furthermore, the ability of positioning a single quantum emitter at the hotspot of the antenna was developed which enables fabrication of highly directional room-temperature single photon sources \cite{Harats_2017,Nikolay_2018, Abudayyeh2020_1, Abudayyeh2020_2}. Another approach of positioning nanodiamonds (NDs) was utilized for a directional emission of single, negatively-charged Nitrogen-Vacancy (NV) centers in NDs positioned on a hybrid metal-dielectric nano-antenna \cite{Nikolay_2018}. A hybrid plasmonic bullseye antenna design based on high-index TiO$_2$ bullseye grating on low-index SiO$_2$ on top a planar silver film has been proposed and validated by improved collection efficiency from NV center in nanodiamonds with reduced background emission as compared to nanostructures of silver films \cite{Andersen_2018}. Also integrated designs have been realized with all-diamond circular bullseye antenna with efficient, broadband collection from single NV center \cite{Li_2015}. 

\begin{figure*}
\includegraphics{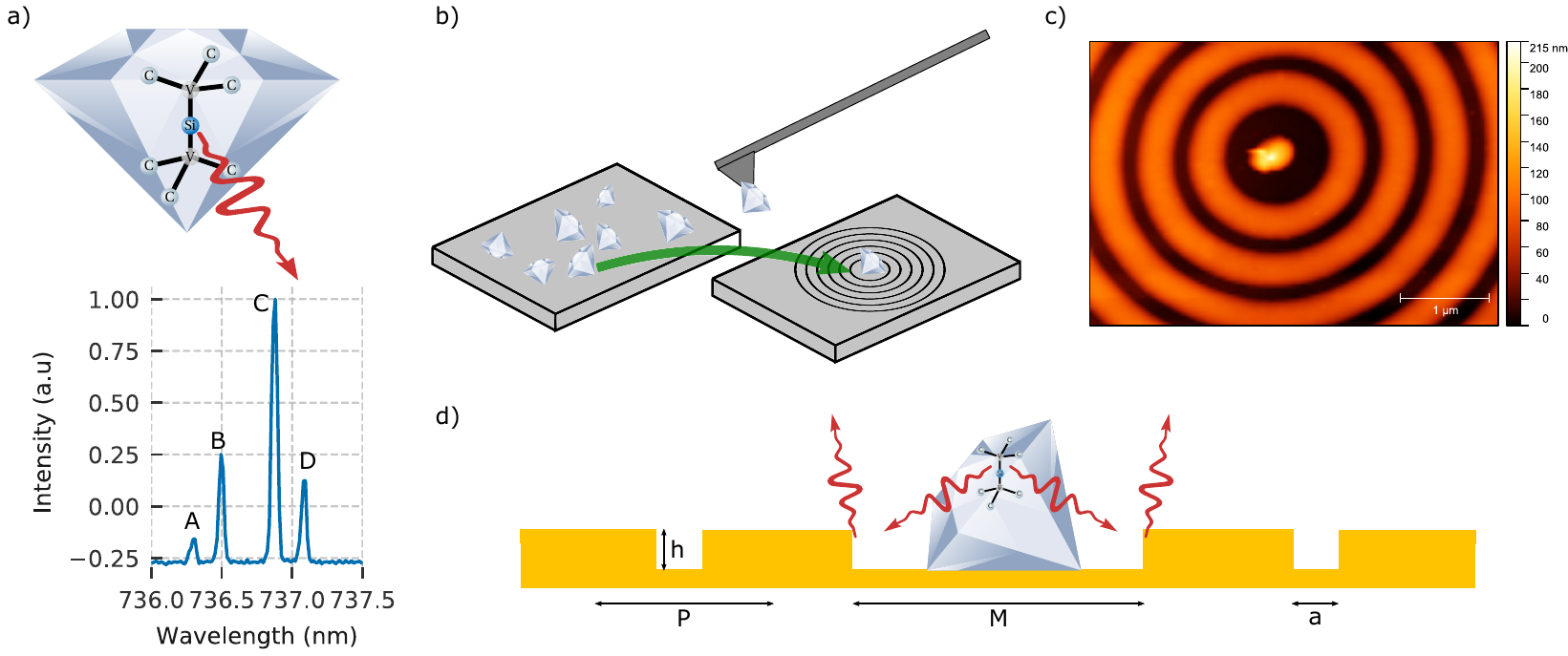}
\caption{\label{fig:fig1}a) Sketch of a nanodiamond hosting a SiV$^-$ color center. Spectrum of a single defect with its characteristic transitions A, B, C, D resulting from the level structure.  b) Illustration of the pick and place method. The nanodiamond is picked from a substrate with the cantilever tip of an AFM and then accurately placed to the center of the bullseye structure. c) AFM scan of the bullseye structure with the placed ND in the center. d) Illustration of the cross section of the ND - bullseye system. The bullseye is formed by rings of gold with a slit width a = 120 nm, and a period P = 450 nm. The height of the rings is h = 80 nm. The ND is placed in the center ring which has a diameter M of 1.375 $\mu$m. Photons are scattered upwards from the bullseye structure.}
\end{figure*} 

In this work, we operate at cryogenic temperatures and focus on efficient light-matter interaction with an inherently coherent solid-state quantum emitter, namely the negatively-charged Silicon-Vacancy (SiV$^-$) center in ND \cite{Rogers_2019}. We demonstrate increased efficiency of resonant excitation yielding high purity single photon emission. With off-resonant and near-resonant excitation we demonstrate enhanced collection efficiency of coherent photons from the zero phonon line (ZPL) of SiV$^-$ center in NDs. We in particular show that resonant excitation of the SiV$^-$ center is possible even with low NA optics and that the system is robust against misalignment.  

In free-space the SiV$^-$-center behaves like a point source and emits photons into all directions. For instance, an objective with a NA of 0.95 can collect emission in a total cone of around 143.6 degrees, ultimately limiting the amount of collected photons. In addition, a SiV$^-$-center-containing ND placed on a substrate will inevitably emit most of its photons into the substrate of higher refractive index \cite{Lukosz:77}, further decreasing the yield of coherent photons. By placing the SiV$^-$-center-containing ND on a bullseye antenna the otherwise lost emission is coherently directed upwards at every metal ring, therefore increasing the amount of detectable coherent photons. A detailed description of the working principle is found in \cite{Abudayyeh_2017} and \citep{Harats}. This enables to map the emission of the whole back-focal plane of the objective onto a CCD camera \cite{Andersen_2018}. \\
In this work, we remain in the standard operation of our confocal setup which is optimized for an objective with high NA of 0.95 in order to resolve single SiV center. In order to demonstrate the directional effect of the bullseye antenna we then change the objective to moderate NA of 0.5 and low NA of 0.25. The extracted total emission cone of the bullseye antenna of $76 \pm 8$ degrees is closest to the 0.5-NA objective with a collection cone of 60 degrees. Furthermore, the focal spot size of the 0.5-NA objective of 1.32 $\mu$m$^2$ matches the area of the first metal ring of the bullseye antenna, enabling efficient interaction with radially propagating surface plasmons at short distance of around 690 nm to the SiV$^-$ center where plasmonic losses are still small. In contrast, the focal spot size of the 0.95-NA objective of 0.37$\mu$m$^2$ is smaller than the area of the first metal ring and also not mode-matched with its emission angle, therefore mostly collecting the free-space emission of the SiV$^-$ center. The focal spot size of the 0.25-NA objective of 5.3$\mu$m$^2$ covers the area of the three first metal rings and, in principle, also interacts with the SiV$^-$ center via the radially propagating surface plasmons. However, the plasmonic channel is more lossy and the optical mode matching is worse. Consequently, we expect the best performance both in terms of resonant excitation efficiency and collection efficiency for the 0.5-NA optics. We map the efficiency with lateral resolution point-by-point and compare it with the free-space emission. By scanning the confocal excitation, and simultaneously the detection spot, we  map out the increased spot size when using the bullseye antenna leading to an increased area from which coherent photons are emitted. 

\section{\label{sec:Methods}Methods}

Our hybrid quantum photonics platform is based on precharacterized SiV$^{-}$-containing NDs which are deterministically placed by means of highly accurate AFM-nanomanipulation in a metallic bullseye antenna, as depicted in figure \ref{fig:fig1} a) and b). The NDs are synthesized by high pressure-high temperature (HPHT) treatment of the catalyst
metal-free growth system based on homogeneous mixtures of naphtalene, fluorinated graphite and tetrakis(trimethylsilyl)sylane as the silicon doping component \citep{Agafonov2014}. 
We characterize the SiV$^{-}$-containing NDs with a custom-built confocal microscope with high-NA optics, in particular with a 0.95-NA objective. Therefore, we mount the sample in a flow cryostat and cool to temperatures of about 4 K.  Optical excitation is performed either off-resonantly with a 532 nm laser or resonantly as well as near-resonantly at 708 nm with a tunable Ti:Sa laser. The fluorescence is filtered with a 720 nm long-pass filter and detected by a single photon counting module (SPCM) and by a grating spectrometer. For resonant excitation we use a 750 LP filter and scan the laser frequency while detecting the phonon sideband fluorescence. After free-space characterization we pick up the ND with an AFM  and place it in the interaction zone of a metallic bullseye antenna. The AFM-based nanomanipulation enables later position optimization and dipole rotation \citep{Huler_2019}. The successful transfer into the bullseye structure is shown in figure \ref{fig:fig1} c) with an AFM scan resolving the positioned ND. The bullseye has a total diameter of around 20 $\mu$m, while the rings of the bullseye antenna have a height $h$ = 80 nm, a period of $P = 450$ nm and a slit width $a$ = 120 nm. The center ring has a diamenter of $M$ = 1.375 $\mathrm{\mu}$m. Figure \ref{fig:fig1} d) shows a cross section sketch of the bullseye structure. 

\section{\label{sec:Results}Results\protect\\}

\begin{figure}
\includegraphics{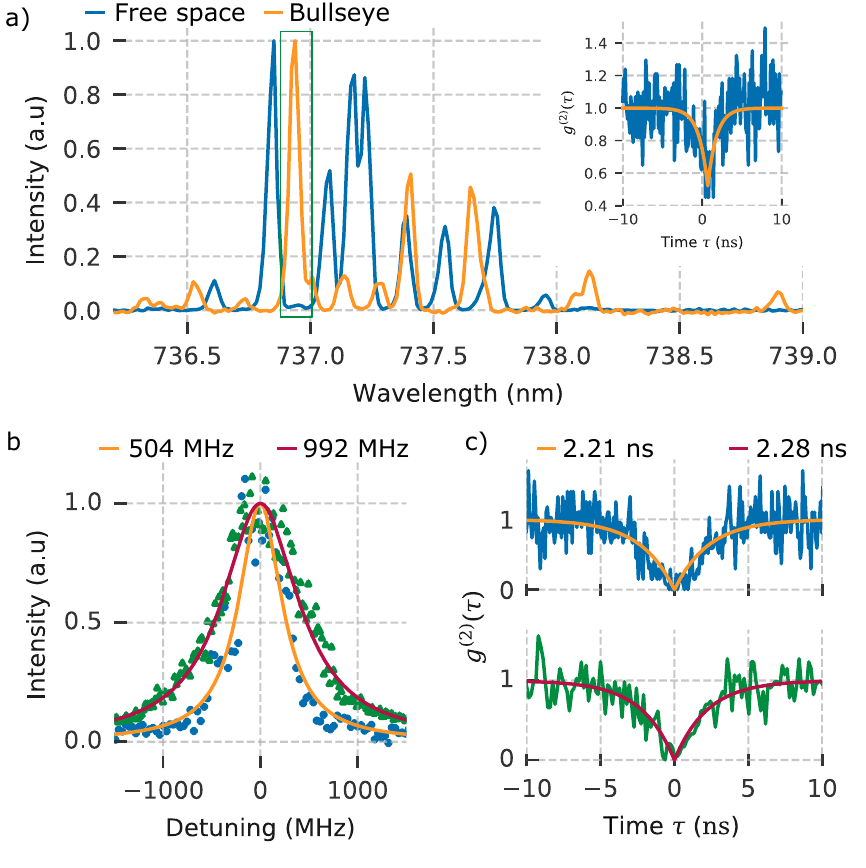}
\caption{\label{fig:fig2}Optical properties of the SiV$^-$ ensemble before and after placing it into the bullseye antenna. a) PL Spectrum in freespace (blue) and from the bullseye antenna(orange). The spectral shift is caused by a change of the strain due to the pick and place method. The marked peak is chosen for the performed measurements. The inset shows an auto-correlation measurement of the marked transition giving a value of $g^2(0) = 0.46$ which indicates two quantum emitters. b) PLE scans of a single transition of the SiV$^-$. Before placing the ND in the bullseye antenna the linewidth is 504 MHz (orange fit). After placing the ND into the bullseye the linewidth is 992 MHz (red fit). c) Resonant second order correlation measurement of the SiV$^-$, both outside (orange) and inside the bullseye antenna (red) clearly indicate single photon emission with a lifetime of (2.21 $\pm$ 0.23) ns and (2.28 $\pm$ 0.21) ns, respectively.}
\end{figure} 

Figure 2 comprises a comparison of the optical properties of the SiV$^-$ center in free-space and after the ND is placed on the bullseye antenna. Since the operation of the bullseye antenna depends on the dipole orientation of the SiV$^-$ center \citep{Abudayyeh_2017} we chose a small cluster of NDs with an overall size of $450 \times 250 \times 200$ nm$^3$ containing about four to six SiV$^-$ center. Assuming a random orientation of the NDs in the cluster this increases the chance to obtain a well-aligned dipole orientation with respect to the bullseye antenna for at least one of the SiV$^-$ center.  Fig. \ref{fig:fig2} a) displays the comparison of the normalized free space emission spectrum (blue) and the spectrum when the ND is placed in the bullseye antenna (orange), both spectra measured at 4 K. The shift in transition frequencies between the spectra most likely originates from modified strain while the differences in relative intensities can be explained by dipole rotation due to the transfer. In order to coherently interact with individual SiV$^-$ center we focus on the most dominant spectral line centered at around 736.74 nm which is highlighted by the green square. Increased spectral resolution uncovers two emission lines centered around 736.720 nm and 736.744 nm, originating from different SiV$^-$ center (see supplementary information). Accordingly, an auto-correlation measurement confirms the presence of two quantum emitters yielding $g^2(0) \approx 0.46$ when both lines are spectrally filtered, as shown in the inset of fig \ref{fig:fig2} a). 

We coherently address the single transition at 736.744 nm by resonant excitation and by detecting the phonon sideband emission. The photoluminescence excitation (PLE) spectroscopy indicates a narrow linewidth of 504 MHz in free-space which increases to 992 MHz after the ND is placed on the bullseye antenna, see fig. \ref{fig:fig2} b). We explain the line broadening by a higher temperature due to decreased thermal contact. Also, the SiV$^-$-center is subject to blinking that occurs on a short time scale without long dark times. A trace is shown in the supplementary information. An auto-correlation measurement under resonant excitation of a single optical transition confirms single photon emission of high purity with sub-poissonian light statistics yielding $g^2(0)=0 \pm 0.066 $, see fig. \ref{fig:fig2} c). Note that no background has been subtracted. The extracted lifetime of (2.28 $\pm$ 0.21) ns of the emission from the bullseye antenna (fig. \ref{fig:fig2} c) lower panel) lies within the error margins of the lifetime of (2.21 $\pm$ 0.23) ns measured in free-space (fig. \ref{fig:fig2} c) upper panel). From auto-correlation measurements with different excitation powers we extrapolate the lifetime at zero power to (2.38 $\pm$ 0.19) ns (see supplementary information).

Next, we perform power-dependent saturation measurements with off-resonant excitation for the regular 0.95-NA objective as well as for a 0.5-NA and a 0.25-NA objective to compare the effect of the bullseye antenna on the excitation efficiency as well as on the collection efficiency. We fit the data with the saturation law (eq. (1)) where $I_{\infty}$ is the saturation count rate and $P_{\mathrm{Sat}}$ the saturation power. 
\begin{equation}
I(\mathrm{P}) = I_{\infty} \frac{P}{P + P_\mathrm{Sat}} 
\end{equation}

\begin{figure}
\includegraphics{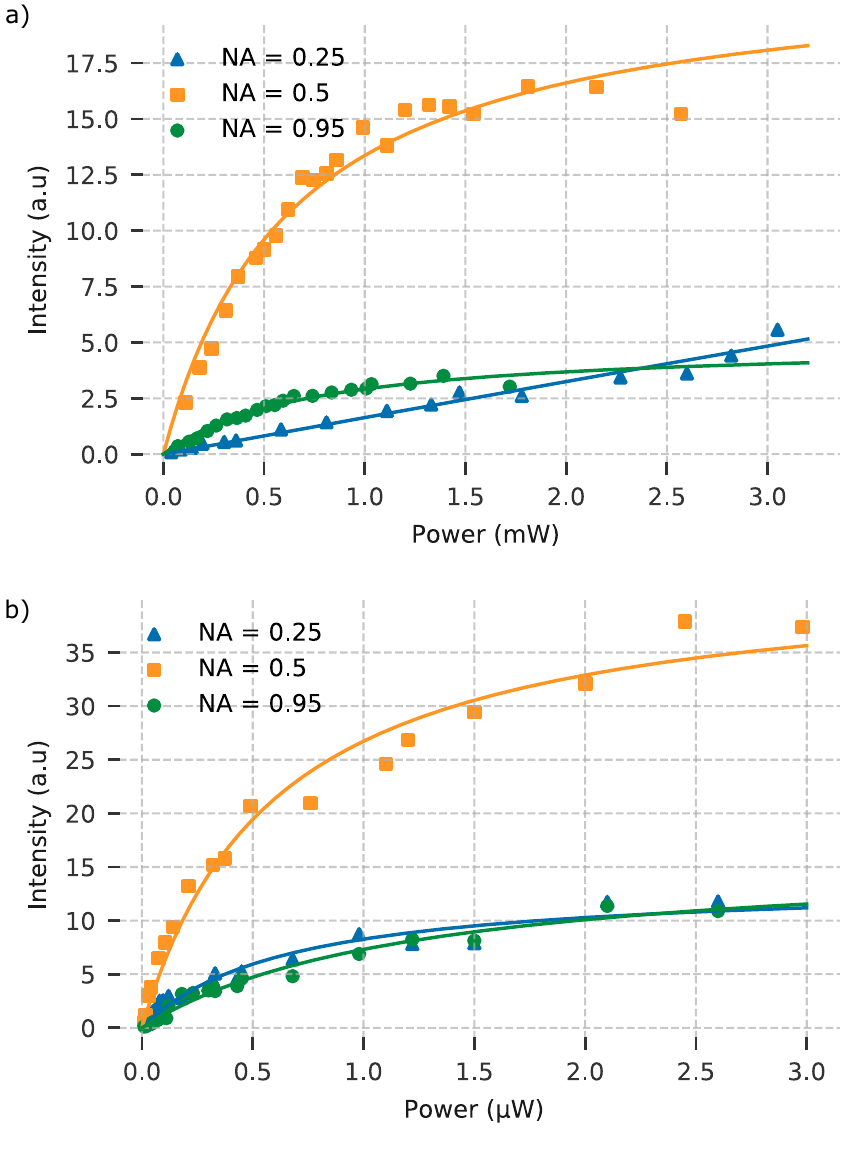}
\caption{\label{fig:fig3}a) Power-dependent measurement of the SiV$^-$ centers for objectives with three different numerical apertures. NA = 0.25 (blue triangle), NA = 0.5 (orange squares) and NA = 0.95 (green dots). The saturation intensity of the ensemble $I_\infty$ measured with the NA of 0.5 is four times higher than when measured with the NA of 0.95. b) Saturation measurement of a single optical transition of the ensemble using resonant excitation, while detecting the phonon sideband fluorescence. The device operates most efficiently with the NA of 0.5.}
\end{figure} 

The saturation curves are shown in fig. \ref{fig:fig3} (a), where blue triangles show data for the 0.25-NA objective, yellow squares for the 0.5-NA objective and green dots for the 0.95-NA objective. We measure a four-fold increase in saturation count rate with the 0.5-NA objective as compared to the 0.95-NA objective. As discussed, the 0.5-NA objective has a better mode-matching with the bullseye antennas emission profile and the focal area is increased by a factor of 3.61 as compared with the 0.95-NA objective. Therefore, while the 0.95-NA objective mostly collects the free-space emission of the SiV$^-$-containing ND placed on the antenna the 0.5-NA objective also collects the fluorescence emitted from the bullseye structure. Here, the fluorescence originates from coupling to radially propagating surface plasmons \citep{Harats} as well as partly-waveguiding, where fluorescence is scattered into the far-field from the antenna. For an ideal dipole, a high-refractive index capping layer would further enhance the waveguiding effect \citep{Abudayyeh_2017}, while here the relatively large ND could take over the waveguiding up to the first metal ring. Note, that the detection efficiency could alter between settings with different objectives. However, the setup was optimized for the 0.95-NA objective and when comparing measurements in free-space the 0.5 NA-objective only gives around 22 \% of the intensity measured with the 0.95-NA objective (see supplementary information).
The saturation power of (0.71 $\pm$ 0.07) mW with the 0.95-NA objective is similar to (0.64 $\pm$ 0.1) mW measured with the 0.5-NA objective, for non-resonant excitation. The saturation power probes the excitation efficiency indicating that under off-resonant excitation there is no significant boost in excitation efficiency originating from the antenna. This observation is in accordance with the fact that the bullseye antenna is optimized for a wavelength of 737 nm and therefore cannot focus the off-resonant excitation of 532 nm efficiently on the SiV$^{-}$ center. In fact, the excitation power density on the SiV$^{-}$ center is reduced with decreasing NA due to an increased spotsize. When operating with low-NA of 0.25 the excitation power density at the position of the SiV$^{-}$ center is reduced so far that saturation is no longer possible with off-resonant excitation. From the different saturation curves we can estimate the ratio of the collection efficiency $\eta_c$ between the 0.95-NA and the 0.5-NA objective. With the relation
\begin{equation}
\frac{\eta_{c1}}{\eta_{c2}} = \frac{P_\mathrm{Sat_2} \cdot I(P_\mathrm{Sat_2})}{P_\mathrm{Sat_1} \cdot I(P_\mathrm{Sat_1})}  
\end{equation}
we find that the collection efficiency of the 0.5-NA objective to be four times the collection efficiency of the 0.95-NA objective. The full-advantage of the bullseye antenna on both the excitation and collection efficiency becomes apparent under resonant drive. Here, the antenna also favors more efficient excitation since the excitation wavelength lies within the operation bandwidth of the device. The bullseye antenna enables efficient resonant excitation with reduced numerical aperture of up to a NA of 0.25. Since the excitation light is now focused efficiently on the SiV$^{-}$ through the bullseye antenna efficient excitation enables saturation of individual transitions even with 0.25-NA optics. The saturation power for the 0.2-NA, 0.5-NA and 0.95-NA is $P_{\mathrm{Sat}}$ = (643 $\pm$ 100) nW, $P_{\mathrm{Sat}}$ = (600 $\pm$ 80) nW and $P_{\mathrm{Sat}}$ = (1219 $\pm$ 160) nW, respectively. Again, we observe highest efficiency with the 0.5-NA objective as summarized in fig. \ref{fig:fig3} b).

\begin{figure}
\includegraphics{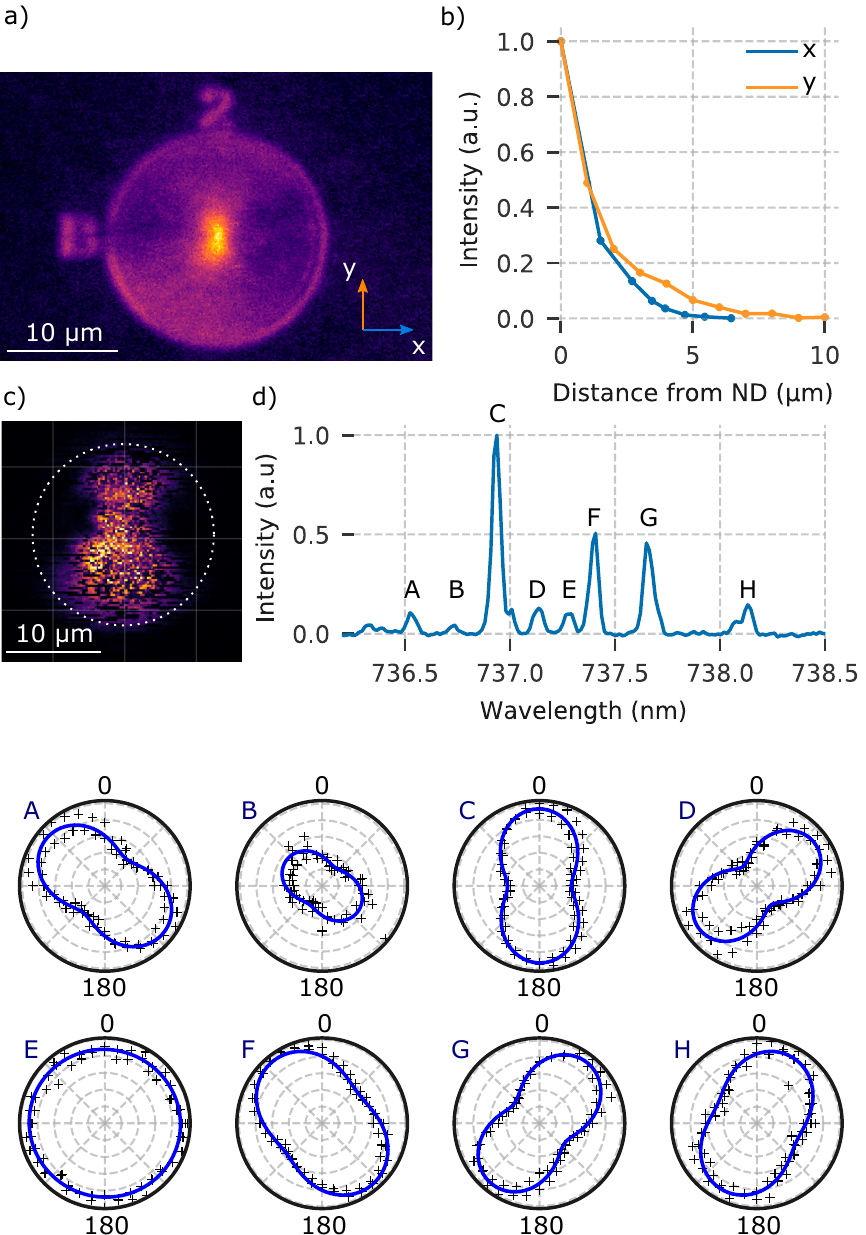}
\caption{\label{fig:fig5}Excitation and collection mapping of the bullseye antenna with the 0.5-NA objective. a) A confocal scan of the antenna, using 708 nm as excitation wavelength. The fluorescence is plotted with respect to the horizontal and vertical axis, marked as x and y. b) The emission of the SiV$^-$-center can be collected up to the edge of the bullseye antenna at a distance of 10 $\mu$m to the ND. At 2 $\mu$m distance, still 50 \% of the emission can be collected. The antenna shows a preferential emission direction, here denoted as y. c) Confocal scan, using resonant excitation of the transition at 736.74 nm. The outline of the bullseye is marked by the dotted circle. Coherent excitation and collection of the emission is possible up to the edge of the bullseye. d) Polarization measurements of visible optical transitions (A to H) of the SiV$^-$ centers. The polarization of the dominant lines marked as C, F and G agrees with the preferential emission direction from the antenna.}
\end{figure}

The bullseye-antenna equipped with SiV$^-$-containing NDs drastically reduces the requirement for sophisticated optical alignment as compared to standard confocal setups. The robustness against optical displacement originates from an increased interaction cross-section and becomes apparent when using moderate or low NA-optics. To map out the interaction area we chose the 0.5-NA optics and a excitation wavelength of 708 nm to scan the excitation spot, and accordingly the detection spot, in confocal configuration. The confocal scan is shown in figure \ref{fig:fig5} a). 50 percent of the signal is still collected when the excitation and collection spot is laterally displaced from the ND with a distance of 2 $\mathrm{\mu}$m. Fig \ref{fig:fig5} b) visualizes the robustness of the bullseye antenna against misalignment. Here, we distinguish two directions, x and y, and record the decreasing fluorescence signal when displacing the confocal spot from the  SiV$^-$-center. The bullseye shows a preferred direction, here the y-direction, where we can still measure fluorescence up to the outer edge of the bullseye antenna. Figure \ref{fig:fig5} c) depicts the collected emission when using resonant excitation (736.74 nm), resulting in a dipole shaped pattern. Here efficient coherent excitation up to the edge of the bullseye is possible. Figure \ref{fig:fig5} d) shows polarization measurements of all visible emission lines (A to H) of the small ensemble of about four to six SiV$^-$ center located in the NDs. The polarization direction of the dominant lines C, F and G is in accordance with the observed preferred emission and pattern in y-direction.

\section{Conclusion}
Summarizing, we demonstrate enhanced detection efficiency for resonant and off-resonant single-photon emission from SiV$^-$-center in NDs placed on a metallic bullseye antenna. The system operates best with moderate-NA optics of about 0.5 and outperforms standard confocal measurements with high NA of 0.95. In addition, under resonant operation the bullseye antenna furthermore increases the excitation efficiency enabling to saturate individual optical transitions even with low-NA optics with a NA of 0.25. The resulting single photon emission is of high purity with $g^2(0)$ close to zero under resonant drive without any background subtraction. The system is robust against misalignment with single photon emission from a large area of almost 100 $\mu m ^2$. Our studied system is therefor very appealing for quantum optical applications, such as distributed quantum information, where coherent light-matter interaction is required and where resonant excitation and efficient single-photon collection is a major requirement. The operation with moderate-NA optics facilitates a robust platform with reduced technical overhead. The efficient operation under off-resonant and near-resonant excitation furthermore enables indistinguishable single photon emission from the zero-phonon line at high rates.
 
The current system relies on enhanced light-matter interaction originating from a combination of coupling to radially propagating surface plasmons and waveguiding due to large index of refraction of the relatively large ND ensemble. The hybrid system is an experimental realization which remains difficult to simulate, taking into account exact size and shape of the NDs as well as the location of the SiV$^-$-center within the NDs. In the future, much smaller NDs on the order of ten nanometers could be placed on the bullseye antenna and capped with a dielectric layer. In such systems, the mechanism of directionality is modified from coupling to radially propagating surface plasmons to coupling to waveguide mode diffraction.

\begin{acknowledgments}
AK acknowledges support of the BMBF/VDI in project Q.Link.X and the European fund for regional development (EFRE) program Baden-Württemberg. AK and RW acknowledges support of the Deutsche Forschungsgemeinschaft (DFG, German Research Foundation) in project 398628099. KGF and AK acknowledge support of IQst. The AFM was funded by the DFG. We thank Prof. Kay Gottschalk and Frederike Erb for their support. VAD thanks the Russian Foundation for Basic Research (grant No. 18-03-00936) for financial support.
\end{acknowledgments}

\providecommand{\noopsort}[1]{}\providecommand{\singleletter}[1]{#1}%

\end{document}